\documentclass [12pt]{article}
\pdfoutput=1

\usepackage{amsmath}
\usepackage{amsfonts}
\usepackage{amscd}
\usepackage{amsthm}
\usepackage{setspace}

\usepackage{graphicx}
\usepackage{authblk}
\usepackage{caption}
\usepackage{ytableau}
\usepackage{mathtools}

\def\Z{\mathbb{Z}}

\def\P{\mathbb{P}}

\begin{document}

\begin{titlepage}

\begin{flushright}
KEK-TH-2201
\end{flushright}

\vskip 3.0cm

\begin{center}

{\bf \Large Types of gauge groups in six-dimensional F-theory on double covers of rational elliptic 3-folds}

\vskip 1.2cm

Yusuke Kimura$^1$ 
\vskip 0.6cm
{\it $^1$KEK Theory Center, Institute of Particle and Nuclear Studies, KEK, \\ 1-1 Oho, Tsukuba, Ibaraki 305-0801, Japan}
\vskip 0.4cm
E-mail: kimurayu@post.kek.jp

\vskip 2cm
\abstract{In this study, we analyze gauge groups in six-dimensional $N=1$ F-theory models. We construct elliptic Calabi--Yau 3-folds possessing various singularity types as double covers of ``1/2 Calabi--Yau 3-folds'', a class of rational elliptic 3-folds, by applying the method discussed in a previous study to classify the singularity types of the 1/2 Calabi--Yau 3-folds. One to three U(1) factors are formed in six-dimensional F-theory on the constructed Calabi--Yau 3-folds. The singularity types of the constructed Calabi--Yau 3-folds corresponding to the non-Abelian gauge group factors in six-dimensional F-theory are deduced. The singularity types of the Calabi--Yau 3-folds constructed in this work consist of $A$- and $D$-type singularities.}  

\end{center}
\end{titlepage}

\tableofcontents
\section{Introduction}
\par U(1) gauge symmetry plays an important role in realizing a grand unified theory (GUT)  since the presence of a U(1) gauge group explains some characteristic properties of a GUT, such as a suppression of the proton decay, and the mass hierarchy of leptons and quarks. The presence of a U(1) gauge group relates to the realization of a GUT in F-theory model building. 
\par F-theory \cite{Vaf, MV1, MV2} provides a framework that extends the Type IIB  superstrings to nonperturbative regimes. F-theory is compactified on spaces that have a torus fibration, wherein the modular parameter of tori as fibers is identified with axio-dilaton, enabling the axio-dilaton to possess an $SL(2,\Z)$ monodromy. 
\par A genus-one fibration is said to have a global section when one can choose a point in each fiber and the chosen point can be moved throughout the base. When an elliptic fibration has a global section, the set of global sections that the fibration possesses forms a group, which is referred to as the ``Mordell--Weil group.'' The number of U(1) factors formed in F-theory, compactified on an elliptic fibration with a section, is equal to the rank of the Mordell--Weil group of that fibration, as discussed in \cite{MV2}. Models of F-theory compactified on elliptic fibrations admitting a global section have been studied, e.g. in \cite{MorrisonPark, MPW, BGK, BMPWsection, CKP, BGK1306, CGKP, CKP1307, CKPS, Mizoguchi1403, AL, EKY1410, LSW, CKPT, CGKPS, MP2, MPT1610, BMW2017, CL2017, KimuraMizoguchi, Kimura1802, LRW2018, TasilectWeigand, MizTani2018, TasilectCL, Kimura1810, CMPV1811, TT2019, Kimura1902, Kimura1903, EJ1905, LW1905, Kimura1910, CKS1910, Kimura1911, FKKMT1912, AFHRZ2001, Kimura2003, KMT2003}. For recent progress in F-theory compactifications where one or more U(1) factors are formed, see, for example, \cite{MorrisonPark, BMPWsection, CKP, CGKP, BMPW2013, CKPS, MTsection, MT2014, KMOPR, BGKintfiber, CKPT, GKK, MPT1610, LS2017, Kimura1802, TT2018, CLLO, CMPV1811, TT2019, Kimura1908, Kimura1910, Kimura1911, OS2019, AFHRZ2001, Kimura2003}. 

\vspace{5mm}

\par The aim of this study is to build six-dimensional (6D) F-theory models with U(1) factors on Calabi--Yau 3-folds, by applying the general scheme developed in \cite{Kimura1910, Kimura2003}, to construct a family of elliptic Calabi--Yau 3-folds of positive Mordell--Weil ranks.  Calabi--Yau 3-folds with $A$-type and $D$-type singularities are obtained in this work, and one to three U(1) gauge group factors are formed in 6D $N=1$ F-theory on the obtained Calabi--Yau 3-folds. We demonstrated by the construction of 6D models that the techniques introduced in \cite{Kimura1910, Kimura2003} apply in studying U(1) gauge symmetries in the F-theory formulation. The singularity types of the constructed Calabi--Yau 3-folds are determined, and they correspond to the non-Abelian gauge groups \cite{MV2, BIKMSV} formed on the 7-branes in 6D F-theory on the constructed Calabi--Yau 3-folds.
\par A certain class of rational elliptic 3-folds was introduced \cite{Kimura1910} to generate elliptic Calabi--Yau 3-folds of various Mordell--Weil ranks, referred to as ``1/2 Calabi--Yau 3-folds.'' Such rational elliptic 3-folds were introduced to build 6D $N=1$ F-theory models with varying numbers of U(1) factors. Because taking double covers of the 1/2 Calabi--Yau 3-folds yield elliptic Calabi--Yau 3-folds \cite{Kimura1910}, 1/2 Calabi--Yau 3-folds can be regarded as ``building blocks'' of Calabi--Yau 3-folds. 
\par Up to seven U(1) factors form in 6D F-theory on the Calabi--Yau 3-folds built as double covers of 1/2 Calabi--Yau 3-folds \cite{Kimura1910}; explicit examples of 6D F-theory models with U(1) have been constructed \cite{Kimura1910}. 
\par To extract information of the non-Abelian gauge groups and matter spectra formed in 6D F-theory, the singularity types and singular fibers \footnote{The types of the singular fibers of the elliptic surfaces were classified in \cite{Kod1, Kod2}, and methods to determine the types of the singular fibers were discussed in \cite{Ner, Tate}.} of Calabi--Yau 3-folds also need to be analyzed. A method to classify the singularity types of the Calabi--Yau 3-folds as double covers of the 1/2 Calabi--Yau 3-folds was discussed in \cite{Kimura2003}. Because the singularity type of the 1/2 Calabi--Yau 3-fold and that of the Calabi--Yau 3-fold as the double cover are identical \cite{Kimura1910}, it suffices to classify the singularity types of the 1/2 Calabi--Yau 3-folds. In addition to this general method of classifying the singularity types, the singularity types of the 1/2 Calabi--Yau 3-folds of rank seven \footnote{Seven is the highest rank of all the singularity types of the 1/2 Calabi--Yau 3-folds \cite{Kimura1910}.} have also been classified, and all types of the 1/2 Calabi--Yau 3-folds with rank seven singularities were explicitly constructed \cite{Kimura2003}. Six-dimensional $N=1$ F-theory on the Calabi--Yau 3-folds, with rank-seven singularity types as their double covers, was also discussed \cite{Kimura2003}. 
\par On this note, we build Calabi--Yau 3-folds of singularity ranks six and lower as double covers of 1/2 Calabi--Yau 3-folds, discussing 6D F-theory on the resulting Calabi--Yau 3-folds. The U(1) gauge group forms in the resulting 6D F-theory models. The singularity types of the constructed Calabi--Yau 3-folds are determined and the number of U(1) factors formed in 6D F-theory are also deduced. The 6D F-theory models on Calabi--Yau 3-folds built as double covers of 1/2 Calabi--Yau 3-folds constructed in this work are new; the 6D F-theory models explicitly constructed in \cite{Kimura1910, Kimura2003} do not include the models we discuss. Our analysis in this study demonstrates that, to a degree, the methods discussed in \cite{Kimura2003} have applications in studying U(1) gauge groups and non-Abelian gauge groups formed in 6D F-theory models. We also discuss the structures of singular fibers of 1/2 Calabi--Yau 3-folds and Calabi--Yau 3-folds as their double covers. These can be used to investigate the non-Abelian gauge symmetries formed on 7-branes. 

\vspace{5mm}

\par Concretely, we construct the 1/2 Calabi--Yau 3-folds with the singularity types: $D_4A_1^2$, $D_5A_1$, $A_3A_2A_1$, $A_3A_1^3$, $A_5A_1$, $A_1^5$, $A_2A_1^3$, $A_3A_1^2$, $A_1^4$. The ranks of these singularity types range from 4 to 6. As the rank of the singularity type and the Mordell--Weil rank of any 1/2 Calabi--Yau 3-fold add to seven \cite{Kimura1910}, the 1/2 Calabi--Yau 3-folds constructed in this work have Mordell--Weil ranks ranging from 1 to 3. One to three U(1) factors form in 6D F-theory compactifications on the Calabi--Yau 3-folds as their double covers. 

\vspace{5mm}

\par A possible relation of 6D $N=1$ F-theory on the Calabi--Yau 3-folds constructed as double covers of 1/2 Calabi--Yau 3-folds to the swampland conditions was mentioned in \cite{Kimura1910, Kimura2003}. The authors of \cite{Vafa05, AMNV06, OV06} discussed the notion of the swampland, and reviews of recent studies on the swampland criteria are given in \cite{BCV1711, Palti1903}. 
\par Possible combinations of distinct gauge symmetries and matter spectra that can form in 6D quantum gravity theories with $N=1$ supersymmetry were discussed in \cite{KT0910, KMT1008, PT1110, Taylor1104}.

\par The structures of elliptic fibrations of 3-folds were analyzed in \cite{Nak, DG, G}.
\par  An emphasis was placed on local model buildings \cite{DWmodel, BHV1, BHV2, DW} in recent model buildings in F-theory. Global aspects of the compactification geometry, however, need to be analyzed to discuss the issues of gravity and problems pertaining to the early universe, including inflation. In this study, the structures of the elliptic Calabi--Yau 3-folds are studied from a global viewpoint. 

\vspace{5mm}

\par This paper is structured as follows. In section \ref{sec2}, applying the discussion \cite{Kimura2003} of the equivalence of the singularity types of the 1/2 Calabi--Yau 3-folds and the plane quartic curves as a consequence of the method in \cite{Muk, Mukai2008, Mukai2019}, we describe the construction of 1/2 Calabi--Yau 3-folds of various singularity types. Taking double covers of the 1/2 Calabi--Yau 3-folds, we construct elliptic Calabi--Yau 3-folds in section \ref{sec3}. F-theory applications are also discussed. U(1) gauge groups are formed in 6D F-theory compactifications on the constructed Calabi--Yau 3-folds. The structures of the singular fibers are analyzed. We state our concluding remarks in section \ref{sec4}. 

\section{Construction of 1/2 Calabi--Yau 3-folds with various singularity types}
\label{sec2}

\subsection{Method of construction}
\label{sec2.1}
We construct 1/2 Calabi--Yau 3-folds of positive Mordell--Weil ranks with various singularity types. F-theory on the Calabi--Yau 3-folds constructed as their double covers have U(1) factors, as we discuss in section \ref{sec3}. The 1/2 Calabi--Yau 3-folds were constructed as a blow-up of $\P^3$ at the base points of three quadrics, as introduced in \cite{Kimura1910}. Taking the ratio $[Q_1: Q_2: Q_3]$ of the three quadrics $Q_1, Q_2, Q_3$ gives a projection onto the base surface $\P^2$, yielding an elliptic fibration. 
\par
\begin{figure}
\begin{center}
\includegraphics[height=10cm, bb=0 0 960 540]{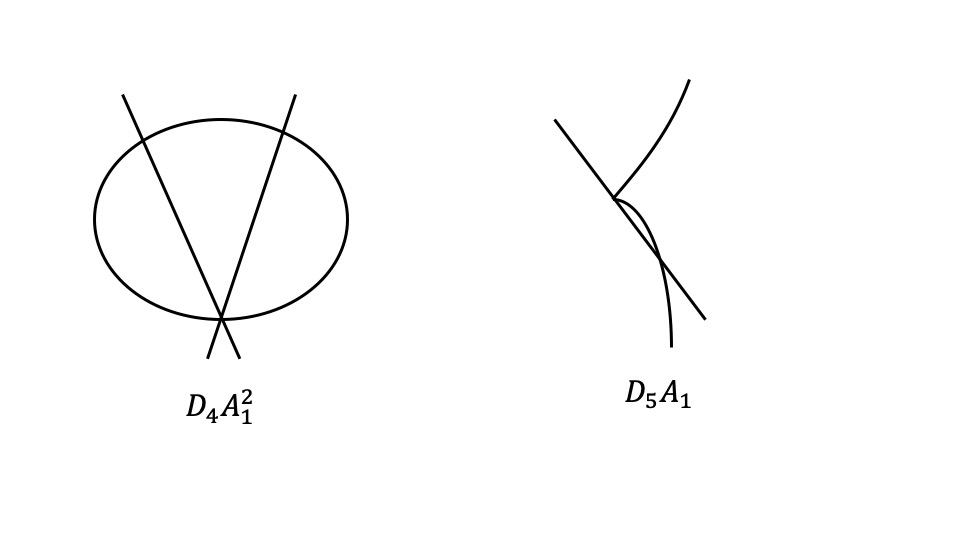}
\caption{\label{imageD4sum2A1} Quartic curve reducible into a conic and two lines intersecting at a common point has a $D_4A_1^2$ singularity (left) \cite{DolgachevAlgGeom}.  The common intersection point yields a $D_4$ singularity, and the other two intersections each yield an $A_1$ singularity. A quartic curve reducible into a cuspidal cubic and a line through the cusp has a $D_5A_1$ singularity (right) \cite{DolgachevAlgGeom}. The cusp yields a $D_5$ singularity, with the other intersection of the line and the cuspidal cubic yielding an $A_1$ singularity.}
\end{center}
\end{figure}

\begin{figure}
\begin{center}
\includegraphics[height=10cm, bb=0 0 960 540]{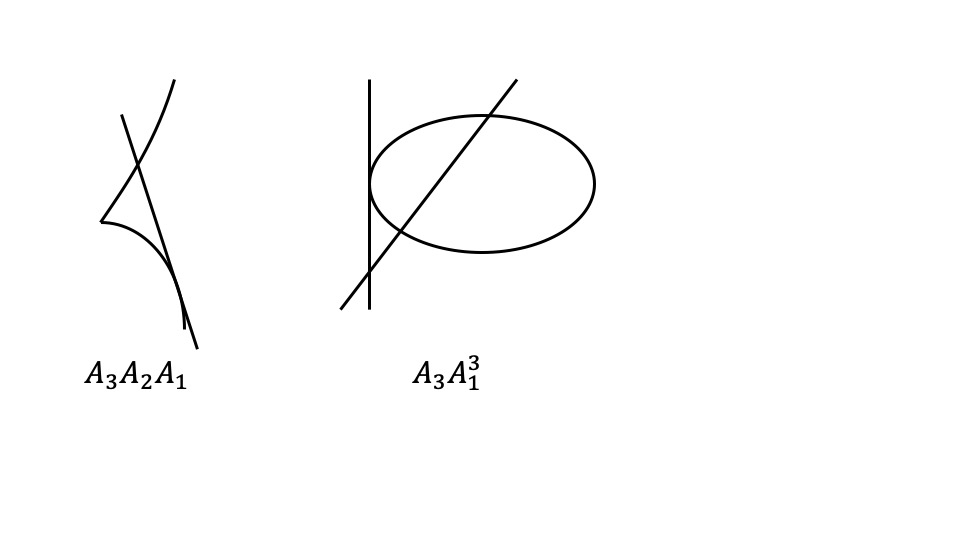}
\caption{\label{imageA3A2A1} Cuspidal cubic and a tangent have an $A_3A_2A_1$ singularity (left) \cite{DolgachevAlgGeom}. The tangent point yields an $A_3$ singularity and the cusp yields an $A_2$ singularity. The other intersection point of the tangent line and the cuspidal cubic yields an $A_1$ singularity. A conic, its tangent, and another line have an $A_3A_1^3$ singularity (right) \cite{DolgachevAlgGeom}. The tangent point yields an $A_3$ singularity, with the other three intersections each yielding an $A_1$ singularity.}
\end{center}
\end{figure}

\begin{figure}
\begin{center}
\includegraphics[height=10cm, bb=0 0 960 540]{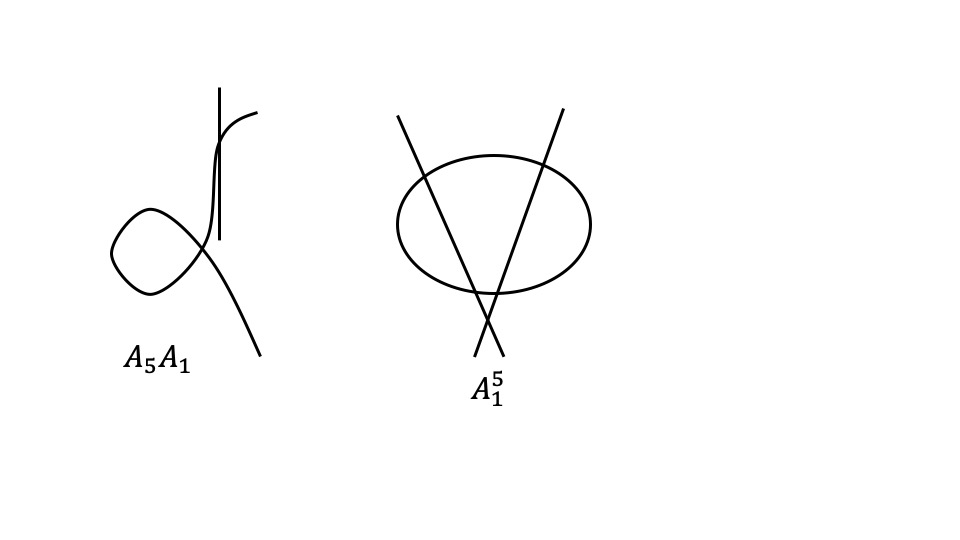}
\caption{\label{imageA5A1} Quartic curve reducible into a nodal cubic and a tangent to flex has an $A_5A_1$ singularity (left) \cite{DolgachevAlgGeom}. The node yields an $A_1$ singularity and the flex tangent yields an $A_5$ singularity. A conic and two lines in a general position have an $A_1^5$ singularity (right) \cite{DolgachevAlgGeom}.} 
\end{center}
\end{figure}

\begin{figure}
\begin{center}
\includegraphics[height=10cm, bb=0 0 960 540]{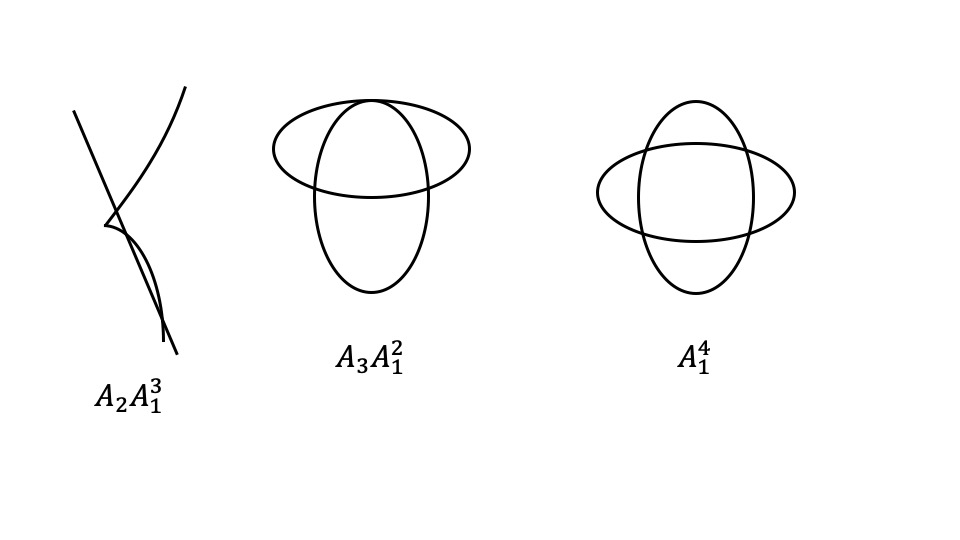}
\caption{\label{imageA2sum3A1} Cuspidal cubic and a line have an $A_2A_1^3$ singularity (left) \cite{DolgachevAlgGeom}. The cusp yields an $A_2$ singularity, with the intersection points of the line and the cuspidal cubic yielding three $A_1$ singularities. Two conics tangent at a point have an $A_3A_1^2$ singularity (middle) \cite{DolgachevAlgGeom}. The tangent point yields an $A_3$ singularity, while the other two intersection points each yield an $A_1$ singularity. Two conics in a general position, intersecting in four distinct points, have an $A_1^4$ singularity (right) \cite{DolgachevAlgGeom}. The four intersections each yield an $A_1$ singularity.}
\end{center}
\end{figure}

 A method to classify the singularity types of the 1/2 Calabi--Yau 3-folds was introduced in \cite{Kimura2003}. Utilizing method discussed in \cite{Muk, Mukai2008, Mukai2019} reveals that the singularity types of the 1/2 Calabi--Yau 3-folds are identical to those of the plane quartic curves \cite{Kimura2003}. The classification of the singularity types of the plane quartic curves is provided in \cite{DolgachevAlgGeom}. We construct 1/2 Calabi--Yau 3-folds with singularity types, $D_4A_1^2$, $D_5A_1$, $A_3A_2A_1$, $A_3A_1^3$, $A_5A_1$, $A_1^5$, $A_2A_1^3$, $A_3A_1^2$, $A_1^4$, by considering the duals of plane quartic curves with identical singularities. The curves with these singularities are shown in Figures \ref{imageD4sum2A1}, \ref{imageA3A2A1}, \ref{imageA5A1}, \ref{imageA2sum3A1}. The equations of the three quadrics yielding 1/2 Calabi--Yau 3-folds with such singularity types are deduced in sections \ref{sec2.2} through \ref{sec2.10}.

\par We used 1/2 Calabi--Yau 3-folds, the construction of which is described in section \ref{sec2.2} through section \ref{sec2.10}, to build elliptic Calabi--Yau 3-folds, discussed in section \ref{sec3}, on which F-theory provides 6D $N=1$ theories.  

\subsection{$D_4A_1^2$ singularity type}
\label{sec2.2}
A quartic curve in $\P^2$ with $D_4A_1^2$ singularity was realized in \cite{DolgachevAlgGeom} as a conic and two lines meeting in a common point. The curve is presented in Figure \ref{imageD4sum2A1} (left). We construct the 1/2 Calabi--Yau 3-fold with $D_4A_1^2$ singularity type as dual of the plane quartic curve. We use $[\lambda:\mu:\nu]$ to denote $\P^2$ into which a quartic curve is embedded. Then the quartic curve with $D_4A_1^2$ singularity is given by the following equation:
\begin{equation}
\label{quartic eqn in 2.2}
(\mu\nu-\lambda^2)\, \lambda\, (\lambda-\mu)=0.
\end{equation}
The common point, $[0:0:1]$, of the conic $\mu\nu-\lambda^2=0$ and the two lines, $\lambda=0$ and $\lambda-\mu=0$, yields $D_4$ singularity. The other intersection point, $[0:1:0]$, of the conic and line $\lambda=0$, and the other intersection, $[1:1:1]$, of the conic and line $\lambda-\mu=0$ yield two $A_1$ singularities. 
\par By an argument similar to that given in \cite{Kimura2003}, the equations of the three quadrics, the blow-up of $\P^3$ at the base points of which yields the 1/2 Calabi--Yau 3-fold with $D_4A_1^2$ singularity type, can be deduced from the determinantal representation of the quartic curve (\ref{quartic eqn in 2.2}). 
\par The determinantal representation of the curve (\ref{quartic eqn in 2.2}) is given as:
\begin{equation}
\label{detrepn in 2.2}
\begin{pmatrix}
\mu & \lambda & 0 & 0 \\
\lambda & \nu & 0 & 0 \\
0 & 0& \lambda & 0 \\
0 & 0& 0 & \lambda-\mu
\end{pmatrix},
\end{equation}
and the equations of the three quadrics $Q_1, Q_2, Q_3$ are deduced from the determinantal representation (\ref{detrepn in 2.2}) as:
\begin{eqnarray}
\label{quadrics D42A1 in 2.2}
Q_1= & z^2+2xy+w^2 \\ \nonumber
Q_2= & x^2-w^2 \\ \nonumber
Q_3= & y^2.
\end{eqnarray}
We used $[x:y:z:w]$ to denote the homogeneous coordinates of $\P^3$. We denote the coordinates of the base surface $\P^2$ of a 1/2 Calabi--Yau 3-fold as $[a:b:c]$. Then, because the curve $c=0$ in the base is dual to the $D_4$ singularity point $[0:0:1]$ of the quartic curve (\ref{quartic eqn in 2.2}), type $I_0^*$ fibers lie over the discriminant curve $c=0$. We use $l_1$ to denote the curve $c=0$. We use $l_2, l_3$ to denote the curves in the base that are dual to the two $A_1$ singularities of the quartic curve (\ref{quartic eqn in 2.2}). Type $I_2$ fibers lie over the curves $l_2$ and $l_3$. 
\par We denote the conic $\mu\nu-\lambda^2=0$ by $C$, and its dual curve as $C^*$ in the base surface of the 1/2 Calabi--Yau 3-fold, then the discriminant of the 1/2 Calabi--Yau 3-fold is given as follows:
\begin{equation}
\Delta \sim l_1^6\cdot l_2^2 \cdot l_3^2\cdot C^*.
\end{equation}

\subsection{$D_5A_1$ singularity type}
\label{sec2.3}
A quartic curve in $\P^2$ with $D_5A_1$ singularity is reducible into a cubic with a cusp \footnote{\cite{Piontkowski} discussed a determinantal representation of the cuspidal cubic.} and a line through the cusp \cite{DolgachevAlgGeom}. We construct the 1/2 Calabi--Yau 3-fold with $D_5A_1$ singularity type as dual of the plane quartic curve. The quartic curve with $D_5A_1$ singularity is given as follows:
\begin{equation}
\label{quartic eqn in 2.3}
(\lambda^3+\mu\nu^2)(\lambda-\nu)=0.
\end{equation}
The cusp is at $[\lambda:\mu:\nu]=[0:1:0]$, and this yields $D_5$ singularity. The other intersection point of the cuspidal cubic $\lambda^3+\mu\nu^2=0$ and the line $\lambda-\nu=0$, $[1:-1:1]$ yields $A_1$ singularity. 
\par The determinantal representation of the quartic curve (\ref{quartic eqn in 2.3}) is given as:
\begin{equation}
\label{detrepn in 2.3}
\begin{pmatrix}
-\mu & 0 & \lambda & 0 \\
0 & -\lambda & \nu & 0 \\
\lambda & \nu& 0 & 0 \\
0 & 0& 0 & \lambda-\nu
\end{pmatrix},
\end{equation}
and the equations of the three quadrics are deduced from the determinantal representation (\ref{detrepn in 2.3}) as:
\begin{eqnarray}
Q_1= & -y^2+2xz+w^2 \\ \nonumber
Q_2= & -x^2 \\ \nonumber
Q_3= & 2yz-w^2.
\end{eqnarray}
\par The curve in the base surface $\P^2$ dual to the $D_5$ singularity point $[0:1:0]$ of the quartic curve (\ref{quartic eqn in 2.3}) is given by $b=0$; We use $l_1$ to denote the curve $b=0$. We denote the curve dual to the $A_1$ singularity of the quartic curve at $[1:-1:1]$ by $l_2$, then type $I_2$ fibers lie over the curve $l_2$. We denote the cuspidal cubic by $B$, and we use $B^*$ to denote its dual in the base surface of the 1/2 Calabi--Yau 3-fold. The discriminant of the 1/2 Calabi--Yau 3-fold with $D_5A_1$ singularity type is given as follows:
\begin{equation}
\Delta \sim l_1^7\cdot l_2^2 \cdot B^*.
\end{equation}

\subsection{$A_3A_2A_1$ singularity type}
\label{sec2.4}
A plane quartic curve with $A_3A_2A_1$ singularity is reducible into a cuspidal cubic and a tangent line \cite{DolgachevAlgGeom}. The curve is presented in Figure \ref{imageA3A2A1} (left). We construct the 1/2 Calabi--Yau 3-fold with $A_3A_2A_1$ singularity type as dual of the plane quartic curve. The quartic curve with $A_3A_2A_1$ singularity is given as follows:
\begin{equation}
\label{quartic eqn in 2.4}
(\lambda^3+\mu\nu^2)(3\lambda+\mu+2\nu)=0.
\end{equation}
The cusp is at $[\lambda:\mu:\nu]=[0:1:0]$ yielding $A_2$ singularity. The point $[\lambda:\mu:\nu]=[1:-1:-1]$ where the line $3\lambda+\mu+2\nu=0$ is tangent to the cuspidal cubic yields $A_3$ singularity. The other intersection point $[2:-8:1]$ of the tangent line and cuspidal cubic yields $A_1$ singularity. 
\par The determinantal representation of the quartic curve (\ref{quartic eqn in 2.4}) is given as:
\begin{equation}
\label{detrepn in 2.4}
\begin{pmatrix}
-\mu & 0 & \lambda & 0 \\
0 & -\lambda & \nu & 0 \\
\lambda & \nu& 0 & 0 \\
0 & 0& 0 & 3\lambda+\mu+2\nu
\end{pmatrix},
\end{equation}
and the equations of the three quadrics are deduced from the determinantal representation (\ref{detrepn in 2.4}) as:
\begin{eqnarray}
\label{quadrics in 2.4}
Q_1= & -y^2+2xz+3w^2 \\ \nonumber
Q_2= & -x^2+w^2 \\ \nonumber
Q_3= & 2yz+2w^2.
\end{eqnarray}
The curve dual to the tangent point is denoted as $l_1$, then the singular fibers over $l_1$ have type $I_4$. The curve in the base surface $\P^2$ dual to the cusp $[0:1:0]$ is given by $b=0$; type $I_3$ fibers lie over this curve. We denote this curve as $l_2$. The curve dual to intersection point $[\lambda:\mu:\nu]=[2:-8:1]$ is denoted as $l_3$, and type $I_2$ fibers lie over the curve $l_3$. The discriminant of the 1/2 Calabi--Yau 3-fold with $A_3A_2A_1$ singularity type is given as follows:
\begin{equation}
\Delta \sim l_1^4\cdot l_2^3\cdot l_3^2 \cdot B^*.
\end{equation}

\subsection{$A_3A_1^3$ singularity type}
\label{sec2.5}
A quartic curve in $\P^2$ with $A_3A_1^3$ singularity is reducible into a conic and a tangent line and another line \cite{DolgachevAlgGeom}. The curve is presented in Figure \ref{imageA3A2A1} (right). We construct the 1/2 Calabi--Yau 3-fold with $A_3A_1^3$ singularity type as dual of the plane quartic curve. The quartic curve with $A_3A_1^3$ singularity is given as follows:
\begin{equation}
\label{quartic eqn in 2.5}
(-\lambda^2+\mu\nu)\, \mu(\mu-\nu)=0.
\end{equation}
The line $\mu=0$ is tangent to the conic $\mu\nu-\lambda^2=0$ at $[0:0:1]$, yielding $A_3$ singularity. Another line $\mu-\nu=0$ intersects with the conic in two points, $[1:1:1], [1:-1:-1]$, and these yield two $A_1$ singularities. Two lines $\mu=0$ and $\mu-\nu=0$ intersect in point $[1:0:0]$ yielding an $A_1$ singularity.
\par The determinantal representation of the quartic curve (\ref{quartic eqn in 2.5}) is given as:
\begin{equation}
\label{detrepn in 2.5}
\begin{pmatrix}
\mu & \lambda & 0 & 0 \\
\lambda & \nu & 0 & 0 \\
0& 0 & \mu & 0 \\
0 & 0& 0 & \mu-\nu
\end{pmatrix},
\end{equation}
and the equations of the three quadrics are deduced from the determinantal representation (\ref{detrepn in 2.5}) as:
\begin{eqnarray}
Q_1= & 2xy \\ \nonumber
Q_2= & x^2+z^2+w^2 \\ \nonumber
Q_3= & y^2-w^2.
\end{eqnarray}
The curve $c=0$ in the base $\P^2$ dual to the tangent point $[0:0:1]$ is denoted as $l_1$, then the singular fibers over the curve $l_1$ have type $I_4$. The curves in the base surface $\P^2$ dual to the three $A_1$ singularities are denoted as $l_2, l_3, l_4$, respectively. The conic $\mu\nu-\lambda^2=0$ is denoted as $C$. The discriminant of the 1/2 Calabi--Yau 3-fold with $A_3A_1^3$ singularity type is given as follows:
\begin{equation}
\Delta \sim l_1^4\cdot l_2^2\cdot l_3^2 \cdot l_4^2 \cdot C^*.
\end{equation}

\subsection{$A_5A_1$ singularity type}
\label{sec2.6}
A plane quartic curve with $A_5A_1$ singularity is reducible into a nodal cubic and a tangent line at a flex \cite{DolgachevAlgGeom}. We construct the 1/2 Calabi--Yau 3-fold with $A_5A_1$ singularity type as dual of the plane quartic curve. The quartic curve with $A_5A_1$ singularity is given as follows:
\begin{equation}
\label{quartic eqn in 2.6}
(\lambda^3+\lambda^2\nu-\mu^2\nu)\nu =0.
\end{equation}
The node at $[\lambda:\mu:\nu]=[0:0:1]$ yields $A_1$ singularity. The flex $[\lambda:\mu:\nu]=[0:1:0]$ where the line $\nu=0$ is tangent to the nodal cubic yields $A_5$ singularity.
\par The determinantal representation of the quartic curve (\ref{quartic eqn in 2.6}) is given as:
\begin{equation}
\label{detrepn in 2.6}
\begin{pmatrix}
\nu & 0 & \lambda & 0 \\
0 & -\lambda & \mu & 0 \\
\lambda & \mu& -\lambda & 0 \\
0 & 0& 0 & \nu
\end{pmatrix},
\end{equation}
and the equations of the three quadrics are deduced from the determinantal representation (\ref{detrepn in 2.6}) as:
\begin{eqnarray}
Q_1= & 2xz-z^2-y^2 \\ \nonumber
Q_2= & 2yz \\ \nonumber
Q_3= & x^2+w^2.
\end{eqnarray}

\subsection{$A_1^5$ singularity type}
\label{sec2.7}
A plane quartic curve with $A_1^5$ singularity was realized in \cite{DolgachevAlgGeom} as a conic and two lines in a general position. We construct the 1/2 Calabi--Yau 3-fold with $A_1^5$ singularity type as dual of the plane quartic curve. Then the quartic curve with $A_1^5$ singularity is given by the following equation:
\begin{equation}
\label{quartic eqn in 2.7}
(\lambda^2-\mu\nu)\, \lambda(\mu-\nu)=0.
\end{equation}
The line $\lambda=0$ intersects with the conic $\lambda^2-\mu\nu=0$ in two points, $[0:1:0], [0:0:1]$. The line $\mu-\nu=0$ intersects with the conic in two points, $[1:1:1], [1:-1:-1]$. The two lines intersect in a point $[0:1:1]$. The five intersection points yield five $A_1$ singularities. 
\par The determinantal representation of the curve (\ref{quartic eqn in 2.7}) is given as:
\begin{equation}
\label{detrepn in 2.7}
\begin{pmatrix}
\mu & \lambda& 0 & 0 \\
\lambda & \nu & 0 & 0 \\
0 & 0& \lambda & 0 \\
0 & 0& 0 & \nu-\mu
\end{pmatrix},
\end{equation}
and the equations of the three quadrics $Q_1, Q_2, Q_3$ are deduced from the determinantal representation (\ref{detrepn in 2.7}) as:
\begin{eqnarray}
Q_1= & z^2+2xy \\ \nonumber
Q_2= & x^2-w^2 \\ \nonumber
Q_3= & y^2+w^2.
\end{eqnarray}
\par We denote the five curves in the base $\P^2$ dual to the five intersection points $[0:1:0], [0:0:1], [1:1:1], [1:-1:-1], [0:1:1]$ as $l_1, l_2, l_3, l_4, l_5$, respectively. The conic $\lambda^2-\mu\nu=0$ is denoted as $C$. The singular fibers over each of the five curves, $l_1, \ldots, l_5$, have type $I_2$. Then, the discriminant of the 1/2 Calabi--Yau 3-fold with $A_1^5$ singularity type is given as follows:
\begin{equation}
\Delta \sim l_1^2\cdot l_2^2 \cdot l_3^2\cdot l_4^2 \cdot l_5^2 \cdot C^*.
\end{equation}

\subsection{$A_2A_1^3$ singularity type}
\label{sec2.8}
A plane quartic curve with $A_2A_1^3$ singularity is reducible into a cuspidal cubic and a line in a general position \cite{DolgachevAlgGeom}. The curve is presented in Figure \ref{imageA2sum3A1} (left). We construct the 1/2 Calabi--Yau 3-fold with $A_2A_1^3$ singularity type as dual of the plane quartic curve. The quartic curve with $A_2A_1^3$ singularity is given as follows:
\begin{equation}
\label{quartic eqn in 2.8}
(\lambda^3+\mu\nu^2)(\lambda+\mu)=0.
\end{equation}
The cusp is at $[\lambda:\mu:\nu]=[0:1:0]$ yielding $A_2$ singularity. Line $\lambda+\mu$ and the cuspidal cubic intersect in three points, $[0:0:1], [1:-1:-1], [1:-1:1]$. The three intersection points yield three $A_1$ singularities.
\par The determinantal representation of the quartic curve (\ref{quartic eqn in 2.8}) is given as:
\begin{equation}
\label{detrepn in 2.8}
\begin{pmatrix}
-\mu & 0 & \lambda & 0 \\
0 & -\lambda & \nu & 0 \\
\lambda & \nu& 0 & 0 \\
0 & 0& 0 & \lambda+\mu
\end{pmatrix},
\end{equation}
and the equations of the three quadrics are deduced from the determinantal representation (\ref{detrepn in 2.8}) as:
\begin{eqnarray}
Q_1= & -y^2+2xz+w^2 \\ \nonumber
Q_2= & -x^2+w^2 \\ \nonumber
Q_3= & 2yz.
\end{eqnarray}
Curves in the base $\P^2$ dual to the three intersection points $[0:0:1], [1:-1:-1], [1:-1:1]$ of the quartic (\ref{quartic eqn in 2.8}) are denoted as $l_1, l_2, l_3$, respectively. The curve in the base surface $\P^2$ dual to the cusp $[0:1:0]$ is given by $b=0$, and this dual curve is denotes as $l_4$. Type $I_2$ fibers lie over the curves $l_1, l_2, l_3$, and type $I_3$ fibers lie over curve $l_4$. The cuspidal cubic is denoted as $B$. The discriminant of the 1/2 Calabi--Yau 3-fold with $A_2A_1^3$ singularity type is given as follows:
\begin{equation}
\Delta \sim l_1^2\cdot l_2^2 \cdot l_3^2\cdot l_4^3 \cdot B^*.
\end{equation}

\subsection{$A_3A_1^2$ singularity type}
\label{sec2.9}
A quartic curve in $\P^2$ with $A_3A_1^2$ singularity is reducible into two conics meeting in three points, one of which is a tangent point \cite{DolgachevAlgGeom}. The curve is presented in Figure \ref{imageA2sum3A1} (middle). We construct the 1/2 Calabi--Yau 3-fold with $A_3A_1^2$ singularity type as dual of the plane quartic curve. The quartic curve with $A_3A_1^2$ singularity is given as follows:
\begin{equation}
\label{quartic eqn in 2.9}
\big(\nu(4\mu+\nu)-\lambda^2\big)\, \big(\lambda\nu-(\mu+\nu)^2 \big)=0.
\end{equation}
The tangent point of the two conics, $\nu(4\mu+\nu)-\lambda^2=0$ and $\lambda\nu-(\mu+\nu)^2=0$, is at $[\lambda:\mu:\nu]=[1:0:1]$, and this yields $A_3$ singularity. The other two intersection points yield $A_1$ singularities. 
\par The determinantal representation of the quartic curve (\ref{quartic eqn in 2.9}) is given as:
\begin{equation}
\label{detrepn in 2.9}
\begin{pmatrix}
4\mu+\nu & \lambda & 0 & 0 \\
\lambda & \nu & 0 & 0 \\
0 & 0 & \lambda & \mu+\nu \\
0 & 0 & \mu+\nu & \nu
\end{pmatrix},
\end{equation}
and the equations of the three quadrics are deduced from the determinantal representation (\ref{detrepn in 2.9}) as:
\begin{eqnarray}
Q_1= & 2xy+z^2 \\ \nonumber
Q_2= & 4x^2+2zw \\ \nonumber
Q_3= & x^2+y^2+w^2+2zw.
\end{eqnarray}
\par The curve in the base surface $\P^2$ dual to the $A_3$ singularity point $[1:0:1]$ of the quartic curve (\ref{quartic eqn in 2.9}) is denoted as $l_1$, and type $I_4$ fibers lie over this curve. We denote the curves dual to two $A_1$ singularities of the quartic curve by $l_2$ and $l_3$, then type $I_2$ fibers lie over the curves $l_2, l_3$. We denote the two conics by $C_1$ and $C_2$, and we use $C^*_1$, $C^*_2$ to denote their duals in the base surface of the 1/2 Calabi--Yau 3-fold. The discriminant of the 1/2 Calabi--Yau 3-fold with $A_3A_1^2$ singularity type is given as follows:
\begin{equation}
\Delta \sim l_1^4\cdot l_2^2 \cdot l_3^2 \cdot C^*_1 \cdot C^*_2.
\end{equation}

\subsection{$A_1^4$ singularity type}
\label{sec2.10}
A plane quartic curve with $A_1^4$ singularity is reducible into two conics in a general position \cite{DolgachevAlgGeom}. We construct the 1/2 Calabi--Yau 3-fold with $A_1^4$ singularity type as dual of the plane quartic curve. The quartic curve with $A_1^4$ singularity is given as follows:
\begin{equation}
\label{quartic eqn in 2.10}
(\mu\nu-\lambda^2)(\lambda\nu-\mu^2)=0.
\end{equation}
The two conics $\mu\nu-\lambda^2=0$ and $\lambda\nu-\mu^2=0$ intersect in four points. The four intersection points yield four $A_1$ singularities.
\par The determinantal representation of the quartic curve (\ref{quartic eqn in 2.10}) is given as:
\begin{equation}
\label{detrepn in 2.10}
\begin{pmatrix}
\mu & \lambda & 0 & 0 \\
\lambda & \nu & 0 & 0 \\
0 & 0 & \lambda & \mu \\
0 & 0& \mu & \nu
\end{pmatrix},
\end{equation}
and the equations of the three quadrics are deduced from the determinantal representation (\ref{detrepn in 2.10}) as:
\begin{eqnarray}
Q_1= & 2xy+z^2 \\ \nonumber
Q_2= & x^2+2zw \\ \nonumber
Q_3= & y^2+w^2.
\end{eqnarray}
Curves in the base $\P^2$ dual to the four intersection points of the quartic (\ref{quartic eqn in 2.10}) are denoted as $l_1, l_2, l_3, l_4$, respectively. Type $I_2$ fibers lie over the curves $l_1, l_2, l_3, l_4$. The two conics are denoted as $C_1$ and $C_2$. The discriminant of the 1/2 Calabi--Yau 3-fold with $A_1^4$ singularity type is given as follows:
\begin{equation}
\Delta \sim l_1^2\cdot l_2^2 \cdot l_3^2\cdot l_4^2 \cdot C_1^* \cdot C_2^*.
\end{equation}

\section{6D F-theory models and gauge groups}
\label{sec3}
We discuss applications to 6D F-theory compactifications. Taking double covers of 1/2 Calabi--Yau 3-folds built in sections \ref{sec2.2} - \ref{sec2.10} yields elliptic Calabi--Yau 3-folds \footnote{The double cover must be ramified over a quartic polynomial in the variables of the three quadrics, $Q_1, Q_2, Q_3$, to satisfy the Calabi--Yau condition. Detail of this can be found in \cite{Kimura1910}.}. F-theory compactifications on the resulting Calabi--Yau 3-folds yield 6D $N=1$ theories. The base surface of the Calabi--Yau 3-folds as double covers of 1/2 Calabi--Yau 3-folds is isomorphic to del Pezzo surface of degree two \cite{Kimura1910} \footnote{We choose a convention to call a del Pezzo surface that is obtained by blowing up $\P^2$ at seven points of a general position as del Pezzo surface of degree two.}; therefore, seven tensor fields arise in 6D $N=1$ F-theory on the Calabi--Yau 3-folds as double covers of 1/2 Calabi--Yau 3-folds \cite{Kimura1910}. 
\par The singularity types of 1/2 Calabi--Yau 3-folds constructed in sections \ref{sec2.2} through \ref{sec2.10} and those of the Calabi--Yau 3-folds as their double covers are identical \cite{Kimura1910}, thus they determine the types of the non-Abelian gauge groups \cite{MV2, BIKMSV} formed in 6D F-theory on the Calabi--Yau 3-folds obtained as the double covers. The singularity types corresponding to the non-Abelian gauge groups formed on the 7-branes in 6D F-theory on the Calabi--Yau 3-folds as double covers of 1/2 Calabi--Yau 3-folds constructed in sections \ref{sec2.2} through \ref{sec2.10} are given as follows: $D_4A_1^2$, $D_5A_1$, $A_3A_2A_1$, $A_3A_1^3$, $A_5A_1$, $A_1^5$, $A_2A_1^3$, $A_3A_1^2$, $A_1^4$. 
\par SU(2) gauge group factors form in 6D F-theory on the Calabi--Yau 3-folds as double covers of 1/2 Calabi--Yau 3-folds whose singularity types include $A_1$, as constructed in sections \ref{sec2.2} through \ref{sec2.10}. For example, SU(2)$^3$ forms in F-theory on the Calabi--Yau 3-fold with $A_3A_1^3$ singularity type, as double cover of 1/2 Calabi--Yau 3-fold constructed in section \ref{sec2.5}.
\par The sum of the rank of the singularity type and the Mordell--Weil rank of any 1/2 Calabi--Yau 3-fold is seven \cite{Kimura1910}. The 1/2 Calabi--Yau 3-folds built in sections \ref{sec2.2} through \ref{sec2.6} have singularity types of rank 6, therefore, their Mordell--Weil rank is 1. The 1/2 Calabi--Yau 3-folds built in sections \ref{sec2.7} through \ref{sec2.9} have rank 5 singularity types, therefore, their Mordell--Weil rank is 2. The 1/2 Calabi--Yau 3-fold built in section \ref{sec2.10} has rank 4 singularity, therefore, the Mordell--Weil rank is 3. 
\par The Mordell--Weil rank of Calabi--Yau 3-fold as double cover of an 1/2 Calabi--Yau 3-fold is greater than or equal to the Mordell--Weil rank of the original 1/2 Calabi--Yau 3-fold \cite{Kimura1910}. Calabi--Yau 3-folds constructed by taking double covers of 1/2 Calabi--Yau 3-folds built in sections \ref{sec2.2} - \ref{sec2.6} have Mordell--Weil ranks greater than or equal to 1. 6D $N=1$ F-theory compactifications on these Calabi--Yau 3-folds have at least one U(1) factor. A similar reasoning applied to 6D F-theory on the Calabi--Yau 3-folds constructed by taking double covers of 1/2 Calabi--Yau 3-folds built in sections \ref{sec2.7} - \ref{sec2.9} shows that at least two U(1) factors form in the theories. As per reasoning similar to these, at least three U(1) factors form in 6D F-theory on the Calabi--Yau 3-folds constructed by taking double covers of 1/2 Calabi--Yau 3-folds built in section \ref{sec2.10}. 
\par The structure of type $I_3$ fibers corresponding to $A_2$ singularities can be seen after conducting blow-ups in our constructions in section \ref{sec2}. The singular fiber corresponding to an $A_2$ singularity is given by two conic meeting in two points, where one and only one of the two meeting points is a base point of the three quadrics. When this base point is blown up, restricted to the singular fiber $\P^1$ arises from the blown-up base point. Then the singular fiber consists of three $\P^1$s, each pair of which meet in one point, and the structure of type $I_3$ fiber becomes clear. 
\par We use the 1/2 Calabi--Yau 3-fold constructed in section \ref{sec2.4} as a sample to demonstrate this point. The singular fibers corresponding to $A_2$ singularity are the fibers over the curve $l_2$, which is dual to the $A_2$ singularity point $[0:1:0]$ of the quartic curve (\ref{quartic eqn in 2.4}). Thus, the equation of the singular fibers is given as follows:
\begin{eqnarray}
\label{singular fiber of A2 before blow-up in 3}
-x^2+w^2= & 0 \\ \nonumber
c(-y^2+2xz+3w^2)-a(2yz+2w^2)= & 0,
\end{eqnarray}
where $[a:c]$ parameterizes the curve $l_2$. We will see that the structure of type $I_3$ fiber can be seen after blow-ups.
\par Before blow-ups, from the equation (\ref{singular fiber of A2 before blow-up in 3}) an argument similar to that given in \cite{Kimura2003} one finds two conics intersecting in two points, where conics are contained in the hyperplanes $x-w=0$ and $x+w=0$. The conics intersect along the intersection of the hyperplanes $x-w=0$ and $x+w=0$. Therefore, the intersection points lie along the locus $x=w=0$, and they are the solutions to the following equation:
\begin{equation}
cy^2+2ayz=0.
\end{equation}
Thus, $\{x=y=w=0\}$ and $\{x=w=0, \hspace{1mm} cy+2az=0\}$ yield two intersection points of the two conics. 
\par
\begin{figure}
\begin{center}
\includegraphics[height=10cm, bb=0 0 960 540]{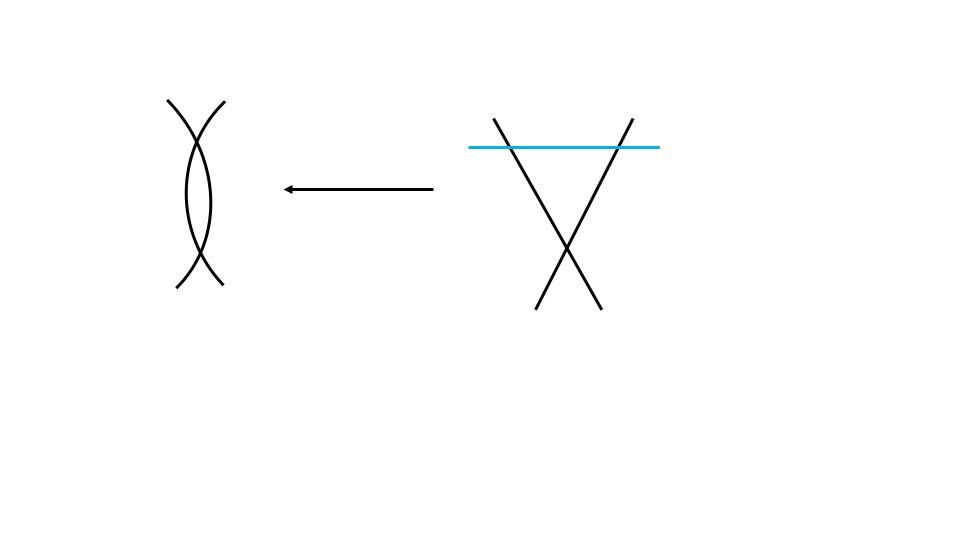}
\caption{\label{imageI3fiber}When one of the intersection points of two conics meeting in two points is blown up, two conics are separated at the blown-up point. $\P^1$ arises from the blown-up point, and the blue line in the right image represents this $\P^1$. This $\P^1$ intersects with each of the two conics in one point, and the structure of type $I_3$ fiber becomes clear as described in the right image.}
\end{center}
\end{figure}

 Three quadrics (\ref{quadrics in 2.4}) have five base points: $[0:0:1:0]$, $[1:1:-1:1]$, $[-1:-1:1:1]$, $[2:-4:1:2]$, $[-2:4:-1:2]$. Therefore, we find that the two intersection points of the conics contain one of the base points: $[0:0:1:0]$. When the base points are blown up, because intersection point $[0:0:1:0]$ is blown up, the blow-up separates the two intersecting conics at the intersection point $[0:0:1:0]$. (The other intersection point remains unchanged.) $\P^1$ arises from the blown-up intersection point, and the structure of type $I_3$ fiber becomes evident after the blow-up. The situation is given in Figure \ref{imageI3fiber}. 

\par The singularity type of 1/2 Calabi--Yau 3-fold constructed in section \ref{sec2.2} includes $D_4$. By an argument similar to that given in \cite{Kimura2003}, we can see the structure of type $I_0^*$ fiber corresponding to $D_4$ singularity after conducting blow-ups at base points of the three quadrics for the 1/2 Calabi--Yau 3-fold in section \ref{sec2.2}. A double conic blown up at four base points yields this fiber type, and the situation of the appearance of a type $I_0^*$ fiber in 1/2 Calabi--Yau 3-fold in section \ref{sec2.2} is analogous to the situation of the 1/2 Calabi--Yau 3-fold with $D_4A_1^3$ singularity discussed in \cite{Kimura2003}. The situations of the $A_3$ singularities of 1/2 Calabi--Yau 3-folds constructed in sections \ref{sec2.4}, \ref{sec2.5}, and \ref{sec2.9} are analogous to the analysis of the 1/2 Calabi--Yau 3-fold with $A_3^2A_1$ singularity studied in \cite{Kimura2003}. Analyses similar to that given in \cite{Kimura2003} also reveal that the structures of type $I_4$ fibers can be seen after the blow-ups of the base points of three quadrics for the constructions in sections \ref{sec2.4}, \ref{sec2.5}, and \ref{sec2.9}, whose singularity types include $A_3$. As mentioned in \cite{Kimura2003}, the structure of type $I_6$ fiber is expected to be seen after multiple stages of blow-ups when a 1/2 Calabi--Yau 3-fold has an $A_5$ singularity as constructed in section \ref{sec2.6}. 
\par The structures of the singular fibers of the 1/2 Calabi--Yau 3-folds are determined via the blow-ups as we have seen. The fiber types of the Calabi--Yau 3-folds as their double covers remain invariant; therefore, the fiber types of the Calabi--Yau 3-folds as the double covers can also be deduced in our constructions. However, whether the types of the singular fibers are split/semi-split/non-split need to be determined to deduce the precise non-Abelian gauge group \cite{BIKMSV}. As pointed out in \cite{Kimura1910}, even when the equations of the three quadrics yielding a 1/2 Calabi--Yau 3-fold are determined, deducing the Weierstrass equation of the Calabi--Yau 3-fold as the double cover from the equations of the three quadrics is technically difficult. Owing to this situation, determining whether a singular fiber is split/non-split/semi-split is generally a hard problem for Calabi--Yau 3-folds constructed as double covers of 1/2 Calabi--Yau 3-folds.
\par The discriminants of the 1/2 Calabi--Yau 3-folds are also deduced in sections \ref{sec2.2}, \ref{sec2.3}, \ref{sec2.4}, \ref{sec2.5}, \ref{sec2.7}, \ref{sec2.8}, \ref{sec2.9}, \ref{sec2.10}. The discriminants of the Calabi--Yau 3-folds constructed as double covers of 1/2 Calabi--Yau 3-folds can be determined from those of the 1/2 Calabi--Yau 3-folds \cite{Kimura1910}. Thus, the locations of the matter fields in 6D F-theory localized at the intersections of the 7-branes can be obtained from the discriminants of the Calabi--Yau 3-folds in our constructions. Determining the matter spectra by studying the collisions of singular fibers at the intersections of the 7-branes can be a likely target of future study. 

\section{Concluding remarks}
\label{sec4}
In this note, we built various elliptic Calabi--Yau 3-folds of positive Mordell--Weil ranks as double covers of 1/2 Calabi--Yau 3-folds of various singularity types via utilizing a general method discussed in \cite{Kimura2003}. F-theory compactifications on the Calabi--Yau 3-folds yielded 6D $N=1$ models wherein one to three U(1) gauge group factors are formed \footnote{Examples of 6D F-theory models on Calabi--Yau 3-folds as double covers of 1/2 Calabi--Yau 3-folds with four to six U(1) factors are constructed in \cite{Kimura1910}. Seven U(1) factors form in 6D F-theory on Calabi--Yau 3-folds constructed as double covers of 1/2 Calabi--Yau 3-folds, when the three quadrics are generically chosen \cite{Kimura1910}.}. 6D F-theory compactifications on the elliptic Calabi--Yau 3-folds with the singularity types, $D_4A_1^2$, $D_5A_1$, $A_3A_2A_1$, $A_3A_1^3$, $A_5A_1$, as double covers of 1/2 Calabi--Yau 3-folds constructed in sections \ref{sec2.2} - \ref{sec2.6}, have at least one U(1) factor; 6D F-theory on Calabi--Yau 3-folds with the singularity types, $A_1^5$, $A_2A_1^3$, $A_3A_1^2$, as double covers of 1/2 Calabi--Yau 3-folds that we constructed in sections \ref{sec2.7} - \ref{sec2.9}, have at least two U(1) factors; and at least three U(1) factors form in 6D F-theory on the elliptic Calabi--Yau 3-folds with the singularity type $A_1^4$ as double cover of 1/2 Calabi--Yau 3-fold as constructed in section \ref{sec2.10}. 
\par Our studies here applied the method in \cite{Kimura2003} to investigate gauge groups formed in 6D F-theory on the Calabi--Yau 3-folds built as the double covers of 1/2 Calabi--Yau 3-folds. A similar analysis can be applied to 1/2 Calabi--Yau 3-folds possessing other singularity types and Calabi--Yau 3-folds as double covers of such 1/2 Calabi--Yau 3-folds. This particularly applies to construct 6D $N=1$ F-theory models with four or more U(1) factors. To analyze such situations, one can construct 1/2 Calabi--Yau 3-folds with singularity types of rank three or less, and consider 6D F-theory on the elliptic Calabi--Yau 3-folds constructed as their double covers. 
\par The structures of the singular fibers are analyzed through the blow-ups as discussed in section \ref{sec3}. There is a chance that the matter spectra can be deduced if the structures of the singular fibers at the intersections of the 7-branes can be analyzed as mentioned in \cite{Kimura1910}, and it might be interesting if this is achieved and the hypermultiplets at the intersections of the 7-branes are determined. This is a likely target for future studies.

\section*{Acknowledgments}

We would like to thank Shigeru Mukai for discussions.

\end{document}